\documentclass[showpacs,aps,graphicx,twocolumn]{revtex4}
\usepackage{graphicx}

\def\ket{\rangle}
\def\<{\langle}
\def\>{\rangle}

\begin{document}

\title{Bidirectional quantum secret sharing and secret splitting with polarized single
photons}
\author{ Fu-Guo Deng,$^{1,2,3,4}$\footnote{fgdeng@bnu.edu.cn}  Hong-Yu Zhou,$ ^{1,2,3}$ and
Gui Lu Long$^{4,5}$\footnote{gllong@tsinghua.edu.cn}}
\address{$^1$ The Key Laboratory of Beam Technology and Material
Modification of Ministry of Education, Beijing Normal University,
Beijing 100875,
China\\
$^2$ Institute of Low Energy
Nuclear Physics, and Department of Material Science and Engineering,
Beijing Normal University, Beijing 100875,  China\\
$^3$ Beijing Radiation Center, Beijing 100875,  China\\
$^4$ Key Laboratory For Quantum Information and Measurements
of Ministry of Education, and Department of Physics, Tsinghua University, Beijing 100084, China\\
$^{5}$ Key Laboratory for Atomic and Molecular Nanosciences,
Tsinghua University, Beijing 100084, China }
\date{\today }

\begin{abstract}
In this Letter, we present quantum secret sharing and secret
splitting protocols with single photons running forth and back
between the participating parties. The protocol  has a high
intrinsic efficiency, namely all photons except those chosen for
eavesdropping check could be used for sharing secret. The
participants need not to announce the measuring bases  at most of
the time and this reduces the classical information exchanged
largely.
\end{abstract}
 \maketitle

The security of  the secret message transmitted has become one of
the most important issues for modern economic and security
activities.  The goal of cryptography is to make secret message
only readable for the two authorized parties,  the sender, Alice
and the receiver, Bob, and unintelligible for any unauthorized
man, say an eavesdropper Eve. To date, the only proven secure
crypto-system in the classical information theory is the
one-time-pad scheme \cite{vernam} in which the key is required to
be as long as the message and is used just one time. The security
of the message transmitted in this scheme depends entirely on the
randomness of the private key.  Alice and Bob have to distribute a
lot of key before they start the secure communication. Quantum key
distribution (QKD) has provided a secure way for transmitting
private keys and it has progressed quickly \cite{Gisin} since an
original QKD protocol was proposed by Bennett and Brassard, the
BB84 scheme \cite{BB84}. The nocloning theorem of quantum state
\cite{nocloning} plays an important role in its security.

Another applications of quantum mechanics within the field of
information is quantum secret sharing (QSS) which is a quantum
counterpart of classical secret sharing \cite{Blakley}. One of the
main goals of QSS is to distribute private keys among the three,
or more generally, multi- parties securely. With the key, the
sender, Alice can divide the message into $N$ shares such that the
other parties can read out the message only when they cooperate,
and any set of less than $N$ shares can get no information about
the message.  An original QSS scheme, the HBB99 scheme
\cite{HBB99} was proposed by Hillery, Bu\v{z}ek and Berthiaume in
1999  using three-particle entangled Greenberger-Horne-Zeilinger
(GHZ) states. In this scheme, the three parties, Alice, Bob and
Charlie choose randomly two MBs, $\sigma_x$ and $\sigma_y$ to
measure the particles in their hands independently. When they all
choose $\sigma_x$ or one chooses $\sigma_x$ and the others choose
$\sigma_y$, their results are correlated and will be kept for
generating key, otherwise they discard the results. Its intrinsic
efficiency, the ratio of number of theoretical valid transmitted
qubits to the number of  transmitted qubits is about 50\% as half
of the instances will be abandoned. Karlsson, Koashi and Imoto
(KKI) put forward a QSS scheme \cite{KKI} with two-photon
polarization-entangled states, and its intrinsic efficiency is
also 50\%. Now, there are many theoretic and experimental studies
on QSS, for instance in Refs.
\cite{cleve,gottesman,Bandyopadhyay,aca,Karimipour,Tyc,
Bagherinezhad,Sen,longqss,delay,Peng,guoqss,TZG}.

Most existing QSS protocols use entangled states and the
participants choose randomly one of two sets of measuring
bases(MBs), for examples the protocols proposed in Refs.
\cite{cleve,gottesman,Bandyopadhyay,aca,Karimipour,Tyc,Bagherinezhad,Sen,longqss,delay,Peng}.
The intrinsic efficiency of these protocols is usually 50\%.  Some
techniques from QKD for improving the intrinsic efficiency
\cite{abc,mbe,delay} can be used for improving their efficiency in
some QSS protocols. For example, the favored-measuring-basis
technique \cite{abc} and the measuring-basis-encrypted technique
\cite{mbe}  are extended to the multiparty QSS schemes in Ref.
\cite{longqss}. In the measuring-basis-encrypted QSS scheme, a
three-party  control key is generated first among the three
parties, and they are used {\it repeatedly} to control the use of
the alternative MBs. With a quantum storage
\cite{storage1,storage2,sun}, delayed measurement is possible and
the efficiency of QSS can also be improved \cite{ delay}.

Entanglement is not necessary in quantum secret sharing. In Ref.
\cite{guoqss}, Guo and Guo proposed a QSS protocol without
entanglement based on a modified BB84 QKD protocol and the
efficiency is improved to approach 100$\%$,  with the use of
quantum data storage. Single photons are ideal source for quantum
communication. However at present, faint laser pulses are used to
as approximate single photon sources. Recently we proposed a QKD
protocol by using  faint laser pulses travelling back and forth
between the two parties \cite{BidQKD}.

In this Letter, we present a  QSS protocol without entanglement by
combining the idea for QSS with QKD \cite{guoqss} and the QKD
protocol with faint laser pulses, the BID-QKD protocol in Ref.
\cite{BidQKD}. In the BID-QKD protocol, photon polarization states
are used to encode information. First Bob prepares photons in one
of the four possible states $|\pm z\ket$, $\pm x\ket$ randomly and
then sends them to Alice. Alice performs some unitary operation to
encode the information and then returns the photons to Bob. Bob
reads out Alice's operation by making measurement. It was shown
that the protocol is still secure if the number of photons in a
pulse does not exceed two. Its intrinsic efficiency is improved to
be 100\%, and it does not require the use of a quantum data
storage. Moreover, the classical information exchanged is reduced
since the parties of communication need not to announce the
information about the measuring basis (MB) in most instances. It
is feasible with present-day technique. We also apply the idea to
quantum secret splitting.

It is noted that any QKD protocol can be used for secret sharing
if Alice can distribute a private key with each of the other
parties. Let us use three parties, Alice, Bob and Charlie in
secret sharing as an example, the naive-QKD QSS protocol. Alice
creates the private key $K_{B}$ with Bob, and $K_{C}$ with
Charlie. However one of the two parties Bob and Charlie maybe
dishonest and the key they use to encrypt are denoted by $K'_{B}$
and $K'_{C}$ respectively. Alice needs to determine whether the
key $K'_A=K'_B\oplus K'_C$ obtained by combining Bob's key and
Charlie's key is the same as her key $K_A=K_B\oplus K_C$, where
$\oplus$ means summing modulo 2. The process can be achieved by
choosing random a sufficiently large subset of bits in the key
$K'_A$ to compare the results with those in the key $K_A$. If the
error rate is zero, Alice confirms that there is no dishonest one
among Bob and Charlie, and she sends the ciphertext to them after
encrypting it with the key $K_A$; otherwise she has to abort the
secret message communication. In this way, secret sharing can be
accomplished with private keys and the main goal of QSS is to
distribute a key among the  parties efficiently.

QSS is more efficient for implementing the task of multi-party
secret sharing than the above naive protocol based on QKD. QSS is
also secure as the legitimate parties can determine eavesdropping.
QSS also reduces the resource requirement \cite{HBB99,KKI,guoqss}.
A figure of merit is the total efficiency $\eta$ defined as
\cite{cabello,longliu}.
\begin{equation}
\eta=\frac{b_s}{q_t+b_t}, \label{eff1}
\end{equation}
where $b_s$ is the number of secret bits in the key, $q_t$ is the
number of qubit used, and $b_t$ is the number of classical bits
exchanged between the parties. For example, the total efficiency
of BB84 \cite{BB84} is $\eta=25\%$ as half of the instances will
be discarded and at least one bit of classical information
exchanged for each qubit, i.e., $b_s=0.5$, $q_t=1$, $b_t=1$. In
the naive-QKD QSS protocol, Alice creates the keys $K_B$ and $K_C$
with Bob and Charlie respectively, and the total efficiency for a
multi-party key is $\eta=\eta_{_B} \cdot \eta_{_c}=12.5\%$. The
HBB99 QSS protocol \cite{HBB99} needs one and half bits of
classical information for each qubit and half  qubit is useful.
Its total efficiency is $\eta_{_{HBB99}}=\frac{0.5}{1+1.5}=20\%$.
So is the KKI QSS protocol \cite{KKI}.

In the following text, we present a QSS protocol using two
bi-directional QKD protocols proposed in Ref. \cite{BidQKD} with
single photons. This protocol  spares the use of quantum data
storage. We will present the idea with a three-party case first,
and the generalization to $N$ parties is also presented.

For creating the key $K_A$, Alice prepares a two-photon product
state $\vert \psi\rangle_A=\vert \phi\rangle_B\otimes \vert
\phi\rangle_C$, and $\vert \phi\rangle_B$ and $\vert
\phi\rangle_C$ are produced with two conjugate bases randomly: the
rectilinear basis $\sigma_z$ (i.e., $\vert +z\rangle=\vert
0\rangle, \,\,\,\, \vert -z\rangle=\vert 1\rangle$) and diagonal
basis $\sigma_x$ (i.e., $ \vert +x\rangle=\frac{1}{\sqrt{2}}(\vert
0\rangle+\vert 1\rangle),
 \,\,\,\, \vert -x\rangle=\frac{1}{\sqrt{2}}(\vert 0\rangle-\vert
 1\rangle)$). Thus the states of the $B$ and $C$ photons at Alice are
randomly in one of the four states $\{\vert +z\rangle, \vert
-z\rangle, \vert +x\rangle, \vert -x\rangle\}$ independently. She
then sends the photon $B$ to Bob and $C$ to Charlie. Bob and
Charlie choose randomly the two unitary operations
$U=i\sigma_y=\vert 0\rangle\langle 1\vert-\vert 1\rangle\langle
0\vert$ and $I=\vert 0\rangle\langle 0\vert+\vert 1\rangle\langle
1\vert$ for most  photons except those chosen randomly for
eavesdropping check. For the sampling photons, they choose
randomly one of the two measuring bases (MBs) $\sigma_z$ and
$\sigma_x$ to measure them, and then they tell Alice their MBs and
results. Alice analyzes the error rate of the sampling photons,
and determines whether there is an eavesdropper in the line. This
is the first eavesdropping check. For other photons, Bob and
Charlie send them back to Alice after encoding with the unitary
operations, and Alice measures them with the same MBs
$\sigma_{_B}\sigma_{_C}\in$\{$\sigma_z\sigma_z$,
$\sigma_z\sigma_x$, $\sigma_x\sigma_z$, $\sigma_x\sigma_x$\} as
she prepares them. The nice feature of the $U$ operation is that
it flips the state in both measuring basis, i.e., the effect of
the operation $U$ is only to negate $($e.g., $\vert 0\rangle$
$\rightarrow$ $\vert 1\rangle$, $\vert 1\rangle$ $\rightarrow$
$\vert 0\rangle$ $)$ the quantum states in the same measuring
basis \cite{QOTP,BidQKD}, i.e.,
\begin{eqnarray}
U\vert +z\rangle&=&-\vert -z\rangle, \,\,\,\, U\vert
-z\rangle=\vert +z\rangle,\\
U\vert +x\rangle&=&\vert -x\rangle, \,\,\,\,  U\vert
-x\rangle=-\vert +x\rangle.
\end{eqnarray}
Alice will get deterministic outcome for each photon returned in
an ideal condition.

For the security of whole process for QSS, Alice needs to choose
randomly a sufficiently subset of result to analyze for the
eavesdropping check after the quantum communication is finished.
This is the second eavesdropping check for creating the
multi-party private key. The difference of the state before she
sends out and receives them is just the combined effect of the
unitary operations performed by Bob and Charlie. As the operations
do not change the photons' MBs, the three parties do not need to
announce the information about the MBs for most photons except the
sampling ones. Then the unitary operations $I$ and $U$ can
represent the bits 0 and 1 respectively, and each photon can carry
one bit of secret message between two parties. The total
efficiency of this QSS approaches $\eta=100\%$ as $b_s=1$,
$b_t=0$, $q_t=1$.

Because some qubits have been used in the two error analysis, the
efficiency is less 1. Suppose there are $\delta$ portion of
transmitted qubits are used in each error checking, then the total
efficiency becomes
\begin{eqnarray}
\eta=(1-\delta)^2,
\end{eqnarray}
and $0<\delta\le 1/2$. In the extreme case, $\delta=1/2$, and the
efficiency becomes 25\%. When $\delta$ approaches zero, $\eta$
approaches 100\%, and in general $0.25\le \eta<1$. Usually
$\delta$ is very small and is negligible, for instance as in
Ref.\cite{lca}, and $\eta$ approaches 1.

We now discuss the security of this QSS. It is pointed out that a
QSS is secure for any eavesdropper if Alice can prevent the
potential dishonest one between Bob and Charlie, say Bob* from
eavesdropping the quantum communication \cite{KKI}. Then the
security depends on the process that Alice and Charlie* can detect
the dishonest one, Bob*, if he eavesdrops the quantum channel. In
fact, the process of this QSS is equal to two BID-QKD protocols
\cite{BidQKD} whose security bases on two BB84 QKD protocols
\cite{BB84,BB84security1,BB84security2,BB84security3} with single
photons.  Alice can synchronously create a private $K_B$ and $K_C$
with Bob and Charlie respectively. In the end, Alice obtains the
key $K_A=K_B\oplus K_C$. For preventing Bob* from eavesdropping,
Alice and Charlie* just accomplish a BID-QKD process which is
secure using single photons as quantum information carrier with
two eavesdropping checks as a unknown state cannot be cloned
\cite{nocloning}, and the action of Bob* will disturb the quantum
system and introduces error in the result in $K_{C*}$. The
relation between the information $I_0(\varepsilon)$ and the error
rate $\varepsilon$ introduced by Bob* can be obtained
\begin{equation}
I_0(\varepsilon)\leq -\varepsilon \log_2
\varepsilon-(1-\varepsilon) \log_2 (1-\varepsilon).\label{pd}
\end{equation}
The probability $P_d=\varepsilon$ that Bob* is detected will
increase with the information $I_0(\varepsilon)$. If the error
rate is low, the information $I_0(\varepsilon)$ is small, and then
the parties can distill a private key with privacy amplification
\cite{Gisin}. Otherwise, they abandon the result. Certainly, the
post-processing should include the error correction part.

It is straightforwardly to generalize this QSS to multi-party
secret sharing. Alice need only prepare a $n$-photon product state
$\vert\psi\rangle_A$ and sends them to the other parties
respectively, i.e., she sends the $i$-th photon to the $i$-th
party. The total wave function Alice prepares is
\begin{equation}
\vert\psi\rangle_A=\prod\limits_{i=1}^{n}\otimes\vert
\phi\rangle_i,
\end{equation}
where $\vert \phi\rangle_i \in \{\vert +z\rangle, \vert -z\rangle,
\vert +x\rangle, \vert -x\rangle\}$. In this way, the multi-party
QSS protocol is composed of $n$ BID-QKD protocols.

Another function of QSS is to split the secret message
\cite{HBB99,KKI}. The present QSS protocol can be used to
accomplish the task if we modify some part of the procedures
following the ideas in quantum secure direct communication
\cite{pingpong,cai2,twostep,QOTP,zhangzj,yanfl}. We also restrict
our discussion to three parties. There are two possible ways for
doing the secret splitting. The first one follows the idea in the
Ping-Pong deterministic secure communication protocol
\cite{pingpong,cai2} in which the photons are transmitted one by
one and it is asymptotically secure when the number of the qubits
transmitted is large \cite{pingpong}. The other one is to use the
quantum secure direct communication (QSDC) protocol \cite{QOTP} in
which the photons are transmitted in a quantum data block and the
message is encoded after the parties of communication confirm that
the quantum channel is secure \cite{QOTP,twostep}. For secret
splitting, Alice prepares a random number string $L$, and adds it
to the secret message $S$, i.e., $G=L\oplus S$. The task of
splitting the message is that Alice sends the string $L$ to Bob
and $G$ to Charlie, and they can read out the message $S=L\oplus
G$ when they collaborate. To this end, Alice requires Bob and
Charlie send to her the polarized photons $B$ and $C$. Assume the
states of photons sent by Bob and Charlie are $\vert
\phi\rangle_B$ and $\vert \phi\rangle_C$ respectively, where
$\vert \phi\rangle_B,\vert \phi\rangle_C \in \{\vert +z\rangle,
\vert -z\rangle, \vert +x\rangle, \vert -x\rangle\}$. Alice
chooses randomly the control mode or the message mode, similar to
Refs. \cite{pingpong} for the photons $B$ and $C$ independently .
When she chooses the control mode, Alice measures the photons one
of the two measuring bases(MBs) , $\sigma_z$ and $\sigma_x$,
randomly. She requires Bob or Charlie to publish her or his
information about the state of the polarized photons. When the
message mode is chosen, Alice encodes $L$ and $C$ on the states
$\vert \phi\rangle_B$ and $\vert \phi\rangle_C$ with the unitary
operation $I$ or $U$ according to the bit 0 or 1 in $L$ and $G$.
She uses the results obtained with control mode as the sampling
photons to analyze the error rate and determines whether there is
an eavesdropper in the line. Alice needs to add some redundancy
randomly on the sequence of $B$ and $C$ photons using the  unitary
operations $I$ and $U$.

In fact, this protocol for splitting the secret message is similar
to two Ping-Pong protocols with single photons \cite{cai2}. The
difference is just that the states of the photons are prepared
with two sets of MBs $\sigma_z$ and $\sigma_x$ randomly as
compared to only a single MB in Ref. \cite{cai2}. Though a small
difference, it improves the security largely. The relation between
the information obtained by an eavesdropper successfully and the
probability that the parties detect her/him is the same as that
for creating a private key, shown in equation (\ref{pd}).

Taking the probability of choosing sampling photons for
eavesdropping check $p_s$ into account, the probability for Bob*
(Eve) to  eavesdrop each qubit successfully is
\begin{equation}
P(1, p_s, \varepsilon)=\frac{1-p_s}{1-(1-\varepsilon)p_s}.
\end{equation}
If Bob* (Eve) eavesdrops $n$ bits of the qubits transmitted, the
probability for Bob* (Eve) to successfully eavesdrop becomes
\begin{equation}
P(n, p_s, \varepsilon)=(\frac{1-p_s}{1-(1-\varepsilon)p_s})^n.
\end{equation}
For $\varepsilon, p_s>0$, the probability $P(n, p_s, \varepsilon)$
decrease exponentially. When the $n$ is sufficiently large, it
approaches zero.

The information that Bob*(Eve) successfully eavesdrops is
$I(\varepsilon)=nI_0(\varepsilon)$ which is small with a low error
rate. For example, if $\varepsilon=0.1, n=10000, P_s=0.1$, then
$I_0(\varepsilon=0.1) \leq 0.47$, $P(10000, p_s,
\varepsilon)=10^{-48}$.

The other way for splitting the secret message can improve the
security with the QSDC protocol \cite{QOTP} based on quantum data
block \cite{QOTP,twostep} at the cost of storing the quantum
states for some times. The whole quantum communication can be
divided into two procedures \cite{QOTP}: (1) the secure doves
sending phase; (2) the message coding and doves returning phase.
It equals to two quantum one time pad QSDC \cite{QOTP}. In the
first phase, Bob and Charlie prepare their quantum state string
$L$ and $G$ independently with the two MBs, $\sigma_z$ and
$\sigma_x$. It means that Bob and Charlie send a group of doves to
Alice respectively. In the second phase, Alice, Bob and Charlie
determine whether there is an eavesdropper or dishonest one
monitoring the quantum channel. Alice encodes the message on the
two groups of the doves if there is no one eavesdropping the
quantum channel, and sends them back to Bob and Charlie.
Otherwise, they abort the communication.

For eavesdropping check, Alice has to store the two sequence of
quantum states that Bob and Charlie prepare with the two MBs
randomly and send to her. The security is discussed in Ref.
\cite{QOTP} in an ideal condition. With a noisy and lossy channel,
the quantum communication can be strengthened  with quantum
privacy amplification \cite{QPA}.

In order to be secure in practice, single photon source and
quantum data storage technique are required. These techniques are
principally available
\cite{singlephoton1,singlephoton2,sun,storage1,storage2}. With the
improvement of technology, the technique may be practically used
for quantum information.

In summary, we have proposed a QSS protocol for creating a private
multi-party key following the idea in bi-directional QKD with
practical faint laser pulse \cite{BidQKD}. The QKD with high total
efficiency is useful for QSS as it reduce the resource requirement
for secret sharing.

This work is supported by the National Natural Science Foundation
of China under Grant Nos. 10435020, 10254002, A0325401, 60433050
and 10325521, the National Fundamental Research Program under
Grant No. 001CB309308, the SRFDP program of Education Ministry of
China.

\end{document}